\documentclass[pra,twocolumn,shownopacs,amssymb]{revtex4}
\usepackage{graphicx}
\usepackage{amsmath}

\begin{document}

\title{Quantum metric contribution to the pair mass in spin-orbit coupled Fermi superfluids}

\author{M. Iskin}
\affiliation{Department of Physics, Ko\c{c} University, Rumelifeneri Yolu, 
34450 Sar\i yer, Istanbul, Turkey}

\date{\today}

\begin{abstract}

As a measure of the quantum distance between Bloch states in the Hilbert space, 
the quantum metric was introduced to solid-state physics through the real part of 
the so-called geometric Fubini-Study tensor, the imaginary part of which 
corresponds to the Berry curvature measuring the emergent gauge field in 
momentum space. Here, we first derive the Ginzburg-Landau theory near the 
critical superfluid transition temperature, and then identify and analyze the 
geometric effects on the effective mass tensor of the Cooper pairs. 
By showing that the quantum metric contribution accounts for a sizeable 
fraction of the pair mass in a surprisingly large parameter regime throughout 
the BCS-BEC crossover, we not only reveal the physical origin of its governing 
role in the superfluid density tensor but also hint at its plausible roles in many 
other observables as well.

\end{abstract}


\maketitle

\section{Introduction}
\label{sec:intro}
Along with the Berry curvature, the quantum metric encodes some of the elusive 
quantum geometrical properties of not only the modern solid-state materials and
condensed-matter systems (e.g., the quantum and spin Hall states, and topological
insulators and superconductors~\cite{niu10, qi11, resta11, sinova15, bansil16})
but also of the photonics systems (e.g., the gyromagnetic photonic crystals, 
coupled resonators and waveguides, bianisotropic metamaterials, and 
quasicrystals~\cite{lu14, wimmer17}), and of the quantum gases 
(e.g., the Hofstadter and Haldane models, and topological and geometrical charge 
pumps~\cite{jotzu14, duca15, aidelsburger15, nakajima16, Lu16}).
Even though both the Berry curvature and the quantum metric characterize by 
definition~\cite{provost80} the local momentum-space geometry of the underlying 
Bloch states, they may also be linked with the global properties of the system 
in somewhat peculiar ways. For instance, the topological Chern invariant of a 
quantum Hall system is simply determined by an integration of the Berry curvature 
over its entire Brillouin zone, controlling the Hall 
conductivity~\cite{niu10, qi11, resta11, sinova15, bansil16}. Likewise, in the context 
of multi-band superconductors, in addition to the conventional intra-band 
contribution determined by the electronic spectra of the Bloch bands, the 
superfluid (SF) weight has an additional contribution coming from the so-called 
geometric inter-band processes~\cite{torma17a}. In the particular 
case of isolated (but not necessarily flat) band superconductors, the inter-band 
contribution is determined by an integration that depends explicitly on the 
quantum metric of the isolated Bloch band, and it is precisely the presence 
of this contribution that prevails superfluidity in the isolated flat-band 
limit for which the intra-band contribution necessarily 
vanishes~\cite{torma15, torma16}.

Despite a long history of inter-disciplinary interest on a variety of physical 
phenomena controlled by the Berry 
curvature~\cite{niu10, qi11, resta11, sinova15, bansil16, lu14, wimmer17, jotzu14, 
duca15, aidelsburger15, nakajima16, Lu16}, nature has not so far
been as generous to the quantum metric effects, given its relatively more 
recent and very limited applications in condensed-matter physics 
theory~\cite{marzari97, zanardi07, haldane11, mudry13, roy14, roy15, 
niu14, niu15, claassen15, atac15, kim15, piechon16, ozawa18, torma15, torma16, 
torma17a, iskin18a, iskin18b}. For instance, the quantum metric of the 
non-interacting helicity bands, which are characterized by the projection of the 
spin onto the direction of momentum in the presence of spin-orbit coupling 
(SOC), is proposed to be responsible for up to a quarter of the total SF
density~\cite{iskin18b}. This is surely a very timely offering motivated 
by the recent creation of Rashba SOC with ultracold Bose and Fermi 
gases~\cite{wu16, sun17, huang16, meng16}. 
Inspired by these developments, here we first identify and then show that 
the quantum metric contribution accounts for a sizeable fraction of the 
effective mass tensor of the Cooper pairs in a surprisingly large parameter 
regime throughout the BCS-BEC crossover. This finding not only reveals the 
physical origin of the quantum metric's governing role in the SF density 
tensor~\cite{torma15, torma16, torma17a, iskin18a, iskin18b} but also hint at its 
plausible roles in many other observables including the sound velocity, 
atomic compressibility, spin susceptibility, etc., all of which depend explicitly 
on the pair mass.
Furthermore, even though most of our analysis is specified for the spin-orbit 
coupled systems, our starting Hamiltonian is quite generic and may find 
direct applications in other two-band SFs as well~\cite{bernevig06, kopnin08}.

\section{Single-particle Hamiltonian and two-body problem}
\label{sec:sp}
Having a two-component or pseudospin-$1/2$ Fermi gas with a generic SOC in mind, 
here we consider a class of single-particle problems that are described by 
the wave equation
$
H_{\mathbf{k}} |s \mathbf{k}\rangle = \epsilon_{s \mathbf{k}} |s \mathbf{k}\rangle,
$
for which the non-interacting Hamiltonian density
$
H_{\mathbf{k}} = \epsilon_\mathbf{k} \sigma_0 + \mathbf{d}_\mathbf{k} \cdot \boldsymbol{\sigma}
$
in $\mathbf{k}$ space leads to a couple of energy (helicity) bands that are 
indexed by $s = \pm$. Thus, the energy spectra of the non-interacting particles 
can be expressed as
$
\epsilon_{s\mathbf{k}} = \epsilon_\mathbf{k} + sd_\mathbf{k},
$
along, respectively, with the following energy eigenstates
$
|s \mathbf{k} \rangle^\textrm{T} = (-d_\mathbf{k}^x+\mathrm{i} d_\mathbf{k}^y, 
d_\mathbf{k}^z - s d_\mathbf{k})/\sqrt{2d_\mathbf{k}(d_\mathbf{k} - s d_\mathbf{k}^z)},
$
where $\textrm{T}$ is the transpose operator.
Here, $\mathbf{k} = \sum_i k_i \boldsymbol{\widehat{i}}
$
is the wave vector, $\epsilon_\mathbf{k} = k^2/(2m)$ with $\hbar \to 1$ is taken 
(in this paper) as the usual quadratic dispersion of a free particle, 
$\sigma_0$ is the $2 \times 2$ identity matrix, 
$
\mathbf{d}_\mathbf{k} = \sum_i d_\mathbf{k}^i \boldsymbol{\widehat{i}} 
$
with the magnitude $d_\mathbf{k} = |\mathbf{d}_\mathbf{k}|$ is the SOC field, and
$
\boldsymbol{\sigma} = \sum_i \sigma_i \boldsymbol{\widehat{i}}
$
is a vector of Pauli spin matrices. Note that $\boldsymbol{\widehat{i}}$ 
is the unit vector along the $\mathbf{i}$ direction in such a way that
$
d_\mathbf{k}^i = \alpha_i k_i 
$
corresponds to a Rashba SOC when $\alpha_x = \alpha_y = \alpha$ and 
$\alpha_z = 0$, and to a Weyl SOC when $\alpha_x = \alpha_y = \alpha_z = \alpha$. 
We choose $\alpha \ge 0$ without losing generality.

%
%

In the presence of an attractive and short-ranged two-body interaction 
between an $\uparrow$ and a $\downarrow$ particle, its strength $U \ge 0$ 
may be linked to the two-body binding energy $\epsilon_b \le 0$ in 
vacuum via the relation,
$
2/U = \sum_{s \mathbf{k}} 1/(2\epsilon_{s \mathbf{k}} + \epsilon_{th} - \epsilon_b),
$
where $\epsilon_{th} = m\alpha^2$ is the energy threshold for the formation
of the two-body bound states. In addition, for a 3D Fermi gas, the theoretical 
parameter $U$ may also be eliminated in favor of the experimentally 
relevant $s$-wave scattering length $a_s$ via the usual relation,
$
1/U = -m V/(4\pi a_s) + \sum_\mathbf{k} 1/(2\epsilon_{\mathbf{k}}),
$
where $V$ is the volume. While the combination of these relations leads to 
an implicit equation of the form
$
1/(m \alpha a_s) = \sqrt{1 + |\epsilon_b|/(m\alpha^2)} 
- \ln ( \sqrt{1+m\alpha^2/|\epsilon_b|}+\sqrt{m\alpha^2/|\epsilon_b|} )
$
for a Rashba SOC, it leads to
$
|\epsilon_b| = 1/(2m a_s^2) + m\alpha^2 \mp \sqrt{1/(4m^2 a_s^4) + \alpha^2/a_s^2}
$
for a Weyl SOC where $\mp$ sign is for the $a_s \lessgtr 0$ region.
Note that setting $|a_s| \to \infty$ in these implicit expressions, we find 
$\epsilon_b \approx -0.44 m\alpha^2$ for the Rashba SOC and 
$\epsilon_b = -m\alpha^2$ for the Weyl SOC at unitarity.
For a 2D Fermi gas, however, eliminating $g$ via the usual relation
$
1/U = \sum_\mathbf{k} 1/(2\epsilon_\mathbf{k} - \epsilon_{sb}),
$
we find
$
|\epsilon_{sb}| = (|\epsilon_b| + m\alpha^2) \exp[-2\sqrt{m\alpha^2/|\epsilon_b|} \arctan(\sqrt{m\alpha^2/|\epsilon_b|})]
$
for a Rashba SOC.

\section{Many-body problem and Ginzburg-Landau theory}
\label{sec:mb}
Once the formation of all sorts of Cooper pairs, e.g., consisting of $\uparrow$ 
particles with $\mathbf{k} + \mathbf{q}/2$ momentum and $\downarrow$ 
particles with $-\mathbf{k} + \mathbf{q}/2$ momentum, is taken into account, 
the resultant effective action may be approximated as
$
S_{eff} \approx S_{0} + S_{Gauss},
$
where the first (second) term is the saddle-point (Gaussian fluctuation) contribution 
coming from the stationary (non-stationary) Cooper pairs with zero (finite) 
center-of-mass momentum $\mathbf{q}$~\cite{zwerger92, melo93, iskin11}. 
Assuming an equal number of particles for the pseudospin components, 
$S_{0} = \Omega_{mf}/T$ is determined by the mean-field thermodynamic 
potential
$
\Omega_{mf} = - T \sum_{s \mathbf{k}} \ln[1 + \exp(-E_{s\mathbf{k}}/T)] 
+ \sum_{s \mathbf{k}} (\xi_{s\mathbf{k}} - E_{s\mathbf{k}})/2
+ \Delta^2/U
$
~\cite{iskin11, jiang11, he12d, he12, shenoy12, ohashi15}, where $T$ is the temperature 
with $k_B \to 1$ the Boltzmann constant,  
$
\xi_{s\mathbf{k}} = \epsilon_{s\mathbf{k}} - \mu
$
is the shifted dispersion for the $s$-helicity band with $\mu$ the chemical 
potential, and
$
E_{s\mathbf{k}} = (\xi_{s\mathbf{k}}^2 + \Delta^2)^{1/2}
$
is the spectrum of the quasiparticles for the corresponding helicity band.
Here, the BCS mean-field
$
\Delta = U \langle \psi_{\uparrow \mathbf{k}} \psi_{\downarrow -\mathbf{k}} \rangle
$
is taken as a real order parameter without losing generality, where 
$\langle \cdots \rangle$ denotes a thermal average over the pair-annihilation 
operator. The mean-field self-consistency equations for $\Delta$ and $\mu$ are
$
1/U = \sum_{s \mathbf{k}} \mathcal{X}_{s \mathbf{k}}/(4E_{s \mathbf{k}})
$
and
$
N_{mf} = \sum_{s \mathbf{k}} [1/2 
- \xi_{s\mathbf{k}} \mathcal{X}_{s\mathbf{k}}/(2E_{s\mathbf{k}})],
$
where
$
\mathcal{X}_{s\mathbf{k}} = \tanh [E_{s\mathbf{k}}/(2T)]
$
is a thermal factor.

%
%

In order to derive the Ginzburg-Landau theory describing the low-energy 
dynamics of the order parameter near the critical SF transition temperature 
$T_c$, we restrict our analysis to its vicinity, and calculate the Gaussian 
contribution $S_{Gauss}$ to the action by expanding the order parameter 
field around $\Delta = 0$ up to quadratic order in the 
fluctuations~\cite{zwerger92, melo93, iskin11}. This leads to
$
S_{Gauss} = (1/T) \sum_{\ell \mathbf{q}} \mathcal{L}_{\ell \mathbf{q}}^{-1}  
|\Lambda_{\ell \mathbf{q}}|^2,
$
where $\Lambda_{\ell \mathbf{q}}$ is the fluctuation field with $\mathbf{q}$ 
the momentum of the Cooper pairs and $\nu_\ell = 2\pi T \ell$ the bosonic 
Matsubara frequency, and
\begin{equation}
\label{eqn:ifp}
\mathcal{L}_{\ell \mathbf{q}}^{-1} = \frac{1}{U} - \frac{1}{8} \sum_{ss'\mathbf{k}} 
\frac{\mathcal{X}_{s +} + \mathcal{X}_{s' -}} {\xi_{s +} + \xi_{s' -} - \mathrm{i}\nu_\ell} 
\left(1 - ss' \widehat{\mathbf{d}}_+ \cdot \widehat{\mathbf{d}}_-\right)
\end{equation}
is the inverse fluctuation propagator~\cite{iskin11, jiang11, he12d, he12, shenoy12, ohashi15}.
Here,
$
\mathcal{X}_{s\pm} = \tanh [\xi_{s\pm}/(2T)]
$
is a short-hand notation with
$
\xi_{s \pm} = \xi_{s, \pm \mathbf{k} + \mathbf{q}/2}
$
for the shifted dispersions, and
$
\widehat{\mathbf{d}}_\pm
$
is for the unit vectors along the SOC fields $\mathbf{d}_{\pm \mathbf{k}+\mathbf{q}/2}$. 
By further expanding the inverse propagator at low-momentum (up to second-order) 
and low-frequency (up to first order in $\omega$ after the analytical continuation
$\textrm{i}\nu_\ell \to \omega + \textrm{i}0^+$), 
\begin{align}
\mathcal{L}_{\omega \mathbf{q}}^{-1} \approx a(T) + \frac{1}{2}\sum_{ij} c_{ij} q_i q_j 
- d_0 \omega + \dots,
\end{align}
we eventually arrive at the celebrated time-dependent Ginzburg-Landau 
equation~\cite{zwerger92, melo93, iskin11}. 
More precisely, the microscopic parameters $a(T)$, $d_0$ and $c_{ij}$ correspond
to the coefficients of its quadratic terms, describing numerous properties of the 
system. We note that the quartic term describes the pair-pair interactions, 
and it is not of particular interest within the scope of this paper. 
For instance, the $T$-dependent coefficient 
$
a(T) = 1/U - \sum_{s \mathbf{k}} \mathcal{X}_{s \mathbf{k}} /(4\xi_{s \mathbf{k}})
$
gives precisely the Thouless criterion $a(T_c) = 0$ for $T_c$, and 
the complex number
$
d_0 = \sum_{s \mathbf{k}} \mathcal{X}_{s \mathbf{k}} / (8\xi_{s \mathbf{k}}^2)
+ \mathrm{i} \pi  \lim_{\omega \to 0^+} \sum_{s \mathbf{k}} \mathcal{X}_{s \mathbf{k}}
\delta(2\xi_{s \mathbf{k}} - \omega) / (4 \xi_{s \mathbf{k}})
$
is the coefficient of the time-dependent term with $\delta(x)$ the Dirac-delta
function. While its nonzero imaginary part for $\mu \ge -m\alpha^2/2$ reflects the 
finite lifetime of the many-body bound states, i.e., due to their instability towards 
decaying into the two-body continuum, its purely real value for $\mu < -m\alpha^2/2$ 
reflects the eventual stability of the two-body bound states that are 
propagating in time with long lifetimes~\cite{melo93, iskin11}.

Most important of all, we notice that the coefficient of the kinetic term 
$c_{ij} = c_{ij}^{intra} + c_{ij}^{inter}$ has two contributions originating from
physically distinct mechanisms, i.e.,
\begin{align}
\label{eqn:intra}
c_{ij}^{intra} &= \sum_{s \mathbf{k}}
\left(
\frac{\mathcal{X}_{s\mathbf{k}}}{16 \xi_{s\mathbf{k}}^2} 
- \frac{\mathcal{Y}_{s\mathbf{k}}}{32T \xi_{s\mathbf{k}}}
\right)
\frac{\partial^2 \xi_{s\mathbf{k}}} {\partial k_i \partial k_j}
\nonumber \\
&\;\;\;\;+ 
\sum_{s \mathbf{k}}
\frac{\mathcal{X}_{s\mathbf{k}} \mathcal{Y}_{s\mathbf{k}}}
{32T^2 \xi_{s\mathbf{k}}}
\frac{\partial \xi_{s\mathbf{k}}} {\partial k_i}
\frac{\partial \xi_{s\mathbf{k}}} {\partial k_j},
\\
c_{ij}^{inter} &= - \sum_{s \mathbf{k}}
\frac{s d_\mathbf{k} \mathcal{X}_{s\mathbf{k}}}{4 \xi_\mathbf{k} \xi_{s\mathbf{k}}} 
g_\mathbf{k}^{ij},
\label{eqn:inter}
\end{align}
where
$
\mathcal{Y}_{s\mathbf{k}} = \mathrm{sech}^2 [\xi_{s\mathbf{k}}/(2T)]
$
is a thermal factor, and
$
g_{\mathbf{k}}^{ij} = (1/2)
\lim_{\mathbf{q} \to \mathbf{0}} 
\partial^2(\widehat{\mathbf{d}}_+ \cdot \widehat{\mathbf{d}}_-)
/ (\partial q_i \partial q_j)
$
or equivalently
$
g_{\mathbf{k}}^{ij} = (\partial \widehat{\mathbf{d}}_\mathbf{k}/\partial k_i) 
\cdot (\partial \widehat{\mathbf{d}}_\mathbf{k}/\partial k_j)/2
$
is solely controlled by the details of the SOC field. 
Here, $c_{ij} = c_{ji}$ is necessarily symmetric in its indices.
The former contribution $c_{ij}^{intra}$ has precisely the conventional form 
arising from the tunneling of the particles within the individual helicity bands, 
and hence its name intra-band. However, the latter contribution $c_{ij}^{inter}$ 
is due to the tunneling of the particles between the helicity bands, and hence 
its name inter-band. Next we show that the inter-helicity contribution has its
roots in the quantum geometry of the underlying $\mathbf{k}$ space, 
making its revelation one of our primary findings in this work.

\section{Quantum metric and Berry curvature}
\label{sec:qm}
First, let us recall that, given a non-interacting multi-band Hamiltonian density 
$H_{\mathbf{k}}$, the quantum metric $g_{n \mathbf{k}}^{ij}$ and the Berry curvature 
$F_{n \mathbf{k}}^{ij}$ of a given Bloch band $n$ are determined by the real and 
imaginary parts of the so-called quantum geometric tensor
$
Q_{n\mathbf{k}}^{ij} = g_{n \mathbf{k}}^{ij} - (\mathrm{i}/2) F_{n \mathbf{k}}^{ij}
$
of the projected Hilbert space defined by
$
(\partial \langle n \mathbf{k} |/\partial k_i) 
(\mathbb{I} - | n \mathbf{k} \rangle \langle n \mathbf{k} |) 
(\partial | n \mathbf{k} \rangle /\partial k_j)
$
~\cite{provost80, resta11}.
Here, the completeness relation is not for the entire Hilbert space, but limited
to the subspace of $\mathbf{k}$ states, in such a way that
$
\mathbb{I} = \sum_{n'} |n' \mathbf{k} \rangle \langle n' \mathbf{k}|
$
with $n'$ summing over all of the available bands.
Alternatively, it is numerically much more practical to implement the elements 
of $Q_{n\mathbf{k}}^{ij}$ tensor in terms of the derivatives of the Hamiltonian 
density as follows
$
Q_{n \mathbf{k}}^{ij} =
\sum_{n' \lbrace \ne n \rbrace} 
\langle n \mathbf{k}| 
(\partial H_{\mathbf{k}}/\partial k_i)
| n' \mathbf{k} \rangle
\langle n' \mathbf{k}| 
(\partial H_{\mathbf{k}}/\partial k_j)
| n \mathbf{k} \rangle
/ (\epsilon_{n \mathbf{k}} - \epsilon_{n' \mathbf{k}})^2.
$
This is because, since the eigenstates $|n \mathbf{k} \rangle$ for a given 
$\mathbf{k}$ are determined up to a random phase factor in a computer program, 
further computation of the derivatives in the original definition produces 
indeterminate factors. Such a numerical ambiguity is clearly avoided by the 
latter formulation. For the case of two bands that are described by our generic 
single-particle problem
$
H_{\mathbf{k}} |s \mathbf{k}\rangle = \epsilon_{s \mathbf{k}} |s \mathbf{k}\rangle,
$
it can be shown analytically that while the quantum metrics
$
g_{s \mathbf{k}}^{ij} = 
(\partial \widehat{\mathbf{d}}_\mathbf{k} / \partial k_i) \cdot 
(\partial \widehat{\mathbf{d}}_\mathbf{k}/ \partial k_j) / 4
$
are identical for both bands, the Berry curvatures are exactly opposite 
$
F_{s \mathbf{k}}^{ij} = sF_{\mathbf{k}}^{ij}
$
of each other with
$
F_{\mathbf{k}}^{ij} = 
[(\partial \widehat{\mathbf{d}}_\mathbf{k} / \partial k_i) 
\times (\partial \widehat{\mathbf{d}}_\mathbf{k} / \partial k_j)]
\cdot \widehat{\mathbf{d}}_\mathbf{k}/2.
$
Thus, while $g_{s \mathbf{k}}^{ij}$ is a symmetric tensor, $F_{s \mathbf{k}}^{ij}$ 
is an anti-symmetric one.
In addition, the components of the latter tensor are determined by the those 
of the former up to a $\mathbf{k}$-dependent sign in the following way
$
|F_{\mathbf{k}}^{ij}|
= (g_{\mathbf{k}}^{ii} g_{\mathbf{k}}^{jj}
- g_{\mathbf{k}}^{ij} g_{\mathbf{k}}^{ji})^{1/2},
$
where
$
g_\mathbf{k}^{ij} = \sum_s g_{s \mathbf{k}}^{ij}
$
is the total quantum metric of the helicity bands appearing explicitly 
in Eq.~(\ref{eqn:inter}).

Given the microscopic coefficients $a(T)$, $d_0$ and $c_{ij}$ of the 
Ginzburg-Landau theory derived above, an effective Gross-Pitaevskii theory 
for the corresponding Bose gas of weakly-interacting pairs can be obtained 
upon the rescaling of the fluctuation field as
$
\Psi_{\omega \mathbf{q}} = \sqrt{d_0} \Lambda_{\omega \mathbf{q}}.
$
Note that this particular choice transforms the coefficient of the time-dependent 
term to that of the Schrodinger one, i.e., it becomes 
$
i \hbar \partial \Psi_{t\mathbf{r}} /\partial t
$ 
in real space $\mathbf{r}$ and time $t$~\cite{melo93, iskin11}.
This identification implies that the effective mass tensor of the Cooper pairs 
is simply given by $m_B^{ij} = d_0/c_{ij}$ for any given set of parameters,
demonstrating the existence of a quantum geometric contribution to the pair mass
in general. As the numerical calculation of these coefficients necessitates 
the self-consistent solutions for $T_c$ and $\mu$ in general, one needs to 
go beyond the mean-field approximation and include the Gaussian 
fluctuation contribution $N_{Gauss}$ to the number equation as the minimal 
prescription for a reliable description. For its simplicity, next we restrict our 
analysis to the weakly-interacting BEC limit of small bosonic molecules, whose 
analytically-tractable nature already illustrates quite convincingly the relative 
importance of the geometric effects without any reliance on heavy numerics.

\section{Fate of Cooper pairs in the molecular Bose gas limit}
\label{sec:ml}
\begin{figure}[htbp]
\includegraphics[scale=0.7]{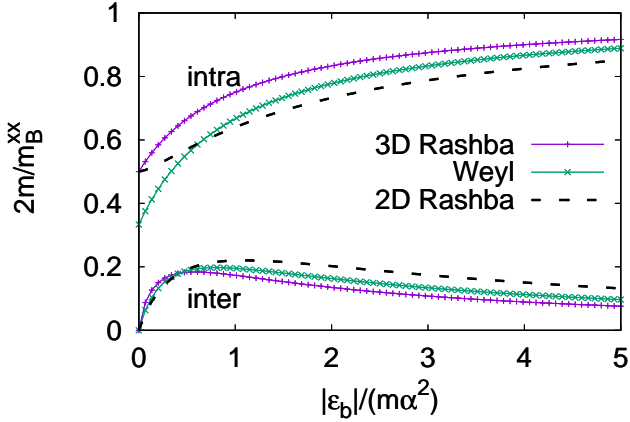}
\caption{(color online)
\label{fig:pairmass}
The intra-helicity and inter-helicity contributions to the effective mass of 
the Cooper pairs $2m/m_B^{xx} = 2m c_{xx}/d_0$ are shown for a molecular 
Bose gas near $T_c$. Since the weakly-interacting BEC limit is characterized by 
$\mu < -m\alpha^2/2$ and $|\mu| = (m\alpha^2 + |\epsilon_b|)/2 \gg T_c$, 
it is possible to achieve this limit by simply increasing $\alpha$ no matter 
how small $|\epsilon_b| \ne 0$ or equivalently interaction strength $U \ne 0$ is.
Note that the peak value of the quantum metric contribution coincides nearly 
with the location of the unitarity in a 3D system, e.g., $|a_s| \to \infty$ when 
$|\epsilon_b| \approx 0.44 m\alpha^2$ for the Rashba SOC and 
$|\epsilon_b| = m\alpha^2$ for the Weyl SOC as discussed in the text.
}
\end{figure}

Since this limit is achieved when $\mu < -m\alpha^2/2$ and $|\mu| \gg T_c$, 
i.e., $\xi_{s \mathbf{k}}/T \to \infty$ for every $\mathbf{k}$, we may simply set 
$\mathcal{X}_{s \mathbf{k}} \to 1$ and $\mathcal{Y}_{s\mathbf{k}} \to 0$ in this limit, 
offering a tremendous simplification of the problem. For instance, eliminating 
$U$ in favor of $\epsilon_b$ and using the Thouless condition $a(T_c) = 0$ with
$
a(T) = 1/U -  \sum_{s \mathbf{k}} 1/(4 \xi_{s\mathbf{k}}),
$
we find $|\mu| = (m\alpha^2 + |\epsilon_b|)/2$ at $T_c$ for the molecular 
Bose gas. In addition, the rest of the coefficients reduce to
$
d_0 = \sum_{s \mathbf{k}} 1/(8 \xi_{s\mathbf{k}}^2)
$
for the time-dependent term, and
$
c_{ij}^{intra} = \sum_{s \mathbf{k}}
[\partial^2 \xi_{s \mathbf{k}}/(\partial k_i \partial k_j)] /(16 \xi_{s\mathbf{k}}^2) 
$
for the intra-helicity contribution and
$
c_{ij}^{inter} = - \sum_{s \mathbf{k}}
s d_\mathbf{k} g_\mathbf{k}^{ij}/(4 \xi_\mathbf{k} \xi_{s\mathbf{k}})
$
for the inter-helicity contribution to the kinetic term.
Note that the derivative 
$
\partial^2 \xi_{s \mathbf{k}}/(\partial k_i \partial k_j)
 = \partial^2 \xi_{\mathbf{k}}/(\partial k_i \partial k_j)
 + s \partial^2 d_{\mathbf{k}}/(\partial k_i \partial k_j)
$
of the helicity spectrum appearing in the intra-helicity contribution may 
also be expressed in terms of the quantum metric as follows
$
\partial^2 d_{\mathbf{k}}/(\partial k_i \partial k_j) 
= 2 d_\mathbf{k} g_\mathbf{k}^{ij} 
+ \widehat{\mathbf{d}}_\mathbf{k} \cdot
\partial^2 \mathbf{d}_{\mathbf{k}}/(\partial k_i \partial k_j).
$
While the latter term vanishes for typical SOCs, the former is in direct competition 
with the inter-helicity contribution due to the difference in their overall signs.
This competition is best seen in Fig.~\ref{fig:pairmass}, where we find that
the intra-helicity (inter-helicity) term has a negative (positive) contribution
to the usual result $2m/m_B = 1$, i.e., when $\alpha \to 0$, in all cases 
considered in this paper.

Prior to presenting our detailed analysis for these coefficients, we note in 
passing that $c_{ij} = c_{ii} \delta_{ij}$, and hence $m_B^{ij}$, is a diagonal 
tensor for the Rashba and Weyl SOCs that are considered in this paper. 
In addition, the critical SF transition temperature $T_c$ of the resultant 
weakly-interacting molecular Bose gas in 3D is well-approximated by the 
critical BEC temperature of a non-interacting Bose gas determined 
by the usual number equation
$
N_B = \sum_\mathbf{q} 1/[\exp(\epsilon_{B \mathbf{q}}/T_c) - 1].
$
Here, $N_B$ is precisely the pole contribution of the Gaussian fluctuations 
characterized by the propagator given in 
Eq.~(\ref{eqn:ifp})~\cite{zwerger92, melo93, iskin11}.
Thus, by plugging
$
\epsilon_{B \mathbf{q}} = \sum_i q_i^2/(2m_B^{ii})
$ 
for the low-energy spectrum of our pairs, we find
$
T_c = 2\pi \{N_B/[V \sqrt{m_B^{xx} m_B^{yy} m_B^{zz}} \zeta(3/2)]\}^{2/3}
$
with $\zeta(3/2) \approx 2.61$ the Riemann-zeta function. Furthermore,
by setting $N_B \approx N/2$ for the pairs with $N = k_F^3 V/(3\pi^2)$ and
the Fermi energy $\epsilon_F = k_F^2/(2m)$, we eventually obtain 
$
T_c/\epsilon_F \approx 0.218 [2m (c_{xx} c_{yy} c_{zz})^{1/3}/d_0]
$
for our molecular Bose gas. In a weakly-interacting molecular 
2D Bose gas, however, $T_c$ is determined by an analogy with the BKT transition,
leading to
$
T_c = \pi N_B/(2Am_B)
$
or equivalently
$
T_c/\epsilon_F = 0.125 (2m c/d_0)
$
for the Rashba SOC, where $N = k_F^2 A/(2\pi)$ with $A$ the area.

\begin{figure}[htbp]
\includegraphics[scale=0.7]{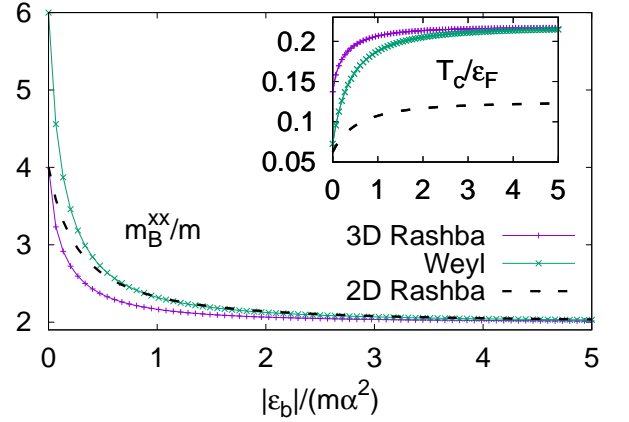}
\caption{(color online)
\label{fig:pairmassandtc}
The effective mass of the Cooper pairs $m_B^{xx} = d_0/c_{xx}$ and the critical 
SF transition temperature $T_c/\epsilon_F$ are shown for a molecular Bose gas. 
It is interesting to see that, since the inter-helicity contribution is a slowly decaying 
function of $|\epsilon_b|/(m\alpha^2)$ as discussed in the main text, it accounts 
for a sizeable fraction of the expected pair mass $m_B = 2m$ even in the 
$|\epsilon_b| \gg m \alpha^2$ limit as long as $\alpha \ne 0$. Thus, we conclude
that the geometric contribution plays an integral role in the proper description 
of the molecular Bose gas. See also~\cite{iskin11, jiang11, he12d, he12, shenoy12, ohashi15}
for similar results.
}
\end{figure}
\subsection{3D Fermi gas with Rashba SOC}
\label{sec:3DR}
In the molecular Bose gas limit, we obtain
$
d_0 = mV\sqrt{2m |\mu|}/[8\pi (2|\mu|-m\alpha^2)]
$
for the time-dependent term, and note that while the kinetic coefficient 
$c_{zz} = c_{zz}^{intra} = d_0/(2m)$ has no inter-helicity contribution, 
its in-plane element $c_{xx} = c_{yy} = c_\perp$ is isotropic in the 
$xy$-plane with the following contribution
$
c_\perp^{intra} = d_0/(2m) - mV\sqrt{2m} \alpha^2 /[64 \pi \sqrt{|\mu|}(2|\mu|-m\alpha^2)]
$
from the intra-helicity component, and
$
c_\perp^{inter} = [V\sqrt{2m}/(64 \pi \sqrt{|\mu|})]
 \ln(2m|\mu|/\sqrt{2m|\mu| - m^2\alpha^2})
$
from the inter-helicity one. Thus, while $m_B^{zz} = 2m$ is purely an 
intra-helicity contribution, we find
$
2m/m_{B\perp}^{intra} = 1 - m\alpha^2/(2|\epsilon_b| + 2m\alpha^2)
$
for the intra-helicity component, and
$
2m/m_{B\perp}^{inter} = [|\epsilon_b|/(2|\epsilon_b| + 2m\alpha^2)] \ln(1 + m\alpha^2/|\epsilon_b|)
$
for the inter-helicity one, which are shown in Fig.~\ref{fig:pairmass}.
Using these analytic results, we conclude that 
$2m/m_{B\perp}^{intra} \to \{1/2, 0.653, 1\}$ and
$2m/m_{B\perp}^{inter} \to \{0, 0.181, 0\}$, 
respectively, when $1/(m \alpha a_s) \to \{-\infty, 0, +\infty\}$,
such that the fraction of the inter-helicity contribution to the pair mass 
is $0.217$ at unitarity.
Note that $T_c/\epsilon_F \to \{0.137, 0.193, 0.218\}$ for the same limits.

\subsection{3D Fermi gas with Weyl SOC}
\label{sec:3DW}
Similar to the Rashba case, here we obtain
$
d_0 = mV\sqrt{m} |\mu|/[4\pi (2|\mu|-m\alpha^2)^{3/2}]
$
for the time-dependent term, and note that the kinetic coefficient 
$c_{xx} = c_{yy} = c_{zz} = c$ is isotropic in all space with the following contribution
$
c^{intra} = d_0/2m - mV\sqrt{m} \alpha^2 /[24\pi (2|\mu|-m\alpha^2)^{3/2}]
$
from the intra-helicity component, and
$
c^{inter} = [V\sqrt{m}/(12\pi)] (1/\sqrt{2|\mu|-m\alpha^2} - 1/\sqrt{2|\mu|})
$
from the inter-helicity one. Thus, we find
$
2m/m_B^{intra} = 1 - 2m\alpha^2/(3|\epsilon_b| + 3m\alpha^2)
$
for the intra-helicity component, and
$
2m/m_B^{inter} = 4|\epsilon_b|/(3|\epsilon_b| + 3m\alpha^2) 
- (4/3)[|\epsilon_b|/(|\epsilon_b| + m\alpha^2)]^{3/2}
$
for the inter-helicity one, which are again shown in Fig.~\ref{fig:pairmass}.
Using these analytic results, we also conclude that 
$2m/m_B^{intra} \to \{1/3, 2/3, 1\}$ and
$2m/m_B^{inter} \to \{0, (2-\sqrt{2})/3, 0\}$, 
respectively, when $1/(m \alpha a_s) \to \{-\infty, 0, +\infty\}$,
such that the fraction of the inter-helicity contribution to the pair mass 
is $0.226$ at unitarity.
Note again that $T_c/\epsilon_F \to \{0.0726, 0.188, 0.218\}$ for the same limits.

\subsection{2D Fermi gas with Rashba SOC}
\label{sec:2DR}
In comparison to the 3D SOCs discussed above, here we obtain
$
d_0 = \lbrace m^3A\alpha/[4\pi(2m |\mu| -m^2 \alpha^2)]\rbrace [1/(m\alpha) + 
\arctan(m \alpha/\sqrt{2m|\mu|-m^2\alpha^2})/\sqrt{2m |\mu| -m^2 \alpha^2}]
$
for the coefficient of the time-dependent term, and note that the kinetic 
coefficient $c_{xx} = c_{yy} = c$ is isotropic in all space with the following 
contribution
$
c^{intra} = d_0/2m - \lbrace m^2 A \alpha /[16\pi(2m |\mu| -m^2 \alpha^2)]\rbrace [\alpha/(2|\mu|) 
+ \arctan(m \alpha/\sqrt{2m|\mu|-m^2\alpha^2})/\sqrt{2m |\mu| -m^2 \alpha^2}]
$
from the intra-helicity component, and a similar contribution
$
c^{inter} =
$
$ 
[m A \alpha /(16\pi |\mu|)]
$
$
\arctan(m \alpha/\sqrt{2m|\mu|-m^2\alpha^2})/
$
$
\sqrt{2m |\mu| -m^2 \alpha^2}]
$
from the inter-helicity component. Thus, we find a lengthy expression
$
2m/m_B^{intra} = 1 - [m\alpha^2/(|\epsilon_b|+m\alpha^2) 
+ \sqrt{m\alpha^2/|\epsilon_b|} \arctan(\sqrt{m\alpha^2/|\epsilon_b|})] /
[2 + 2\sqrt{m\alpha^2/|\epsilon_b|} \arctan(\sqrt{m\alpha^2/|\epsilon_b|})]
$
for the intra-helicity component, and similarly
$
2m/m_B^{inter} = [\sqrt{m\alpha^2 |\epsilon_b|}/(|\epsilon_b| + m\alpha^2) \arctan(\sqrt{m\alpha^2/|\epsilon_b|})] /
[1 + \sqrt{m\alpha^2/|\epsilon_b|} \arctan(\sqrt{m\alpha^2/|\epsilon_b|})]
$
for the inter-helicity component, which are again shown in Fig.~\ref{fig:pairmass}.
Using these analytic results, we conclude that 
$2m/m_B^{intra} \to \{1/2, 1\}$ and $2m/m_B^{inter} \to \{0, 0\}$
in perfect correspondence with the 3D Rashba SOC discussed in case $(I)$, 
respectively, when $|\epsilon_b|/(m \alpha^2) \to \{0, +\infty\}$, and also that 
$T_c/\epsilon_F \to \{0.0625, 0.125\}$ for the same limits. 

In addition to illustrating all of the relevant limits mentioned above, one of the 
noteworthy revelations of Fig.~\ref{fig:pairmass} is that the peak values of the 
quantum metric contribution coincide nearly with the locations of the unitarity 
in 3D systems. More importantly, Fig.~\ref{fig:pairmassandtc} shows that since 
the inter-helicity contribution is a slowly decaying function of $|\epsilon_b|/(m\alpha^2)$, 
it accounts for a sizeable fraction of the expected pair mass $m_B = 2m$ even 
in the $|\epsilon_b| \gg m \alpha^2$ limit as long as $\alpha \ne 0$. 
Thus, our analysis suggests that the quantum metric contribution is a non-trivial
fraction of the pair mass in a fairly large parameter regime in the BCS-BEC 
crossover, making its experimental observation a real possibility with spin-orbit 
coupled Fermi SFs.

\section{Conclusions}
\label{sec:conc}
In summary, here we showed that the quantum metric contribution 
to the effective mass tensor of the Cooper pairs accounts for a sizeable fraction 
of the pair mass in a surprisingly large parameter regime throughout the 
BCS-BEC crossover. This work reveals not only the physical origin of 
the governing role played by the quantum metric in the SF density 
tensor~\cite{torma15, torma16, torma17a, iskin18a, iskin18b} 
but also hint at its plausible roles in many other observables including 
the sound velocity, atomic compressibility, spin susceptibility, etc., 
all of which depend explicitly on the the pair mass. For instance, similar 
to the non-monotonic evolution of the SF density, which is a 
direct consequence of the competition between the intra-helicity and 
inter-helicity contributions at low temperatures~\cite{iskin18b}, we expect 
non-monotonic evolutions for those observables that are proportional to the 
kinetic coefficient $c_{ij}$ or equivalently to the inverse of the pair mass $1/m_B$.  
As a final remark, it is worth pointing out that, even though most of our analysis 
in this paper is restricted to the spin-orbit coupled Fermi SFs, our starting 
Hamiltonian density $H_\mathbf{k}$ is quite generic and may find direct 
applications in other two-band systems as well, e.g., in the contexts of 
quantum spin-Hall effect~\cite{bernevig06} and superconductivity of Dirac 
electrons in graphene layers~\cite{kopnin08}.

\begin{acknowledgments}
The author thanks Atac Imamoglu for encouraging comments, and acknowledges 
support from T{\"U}B{\.I}TAK and the BAGEP award of the Turkish Science Academy.
\end{acknowledgments}

\end{document}